\documentclass[preprint2]{aastex}
\usepackage{graphicx}

\shortauthors{Hopkins et al.}
\shorttitle{Resolving SFR discrepancies}

\begin{document}

\title{Towards a resolution of the discrepancy between different estimators of
star formation rate}

\author{A. M. Hopkins\altaffilmark{1}, A. J. Connolly\altaffilmark{2},} 

\affil{Department of Physics and Astronomy, University of Pittsburgh, 
  3941 O'Hara Street, Pittsburgh, PA 15260, USA}

\and

\author{D. B. Haarsma,} 

\affil{Calvin College, 3201 Burton Street, Grand Rapids, MI 49546, USA}

\and

\author{L. E. Cram}

\affil{School of Physics, University of Sydney, NSW 2006, Australia}

\altaffiltext{1}{Email: ahopkins@phyast.pitt.edu}
\altaffiltext{2}{Email: ajc@phyast.pitt.edu}

\begin{abstract}
Different wavelength regimes and methods for estimating the space
density of the star-formation rate (SFR) result in discrepant values. While
it is recognised that ultra-violet (UV) and H$\alpha$ emission line data must
be corrected for the effects of extinction, the magnitude of the required
correction is uncertain. Even when these corrections are made there remains a
significant discrepancy between SFRs derived from UV and H$\alpha$ measurements
compared with those derived from far-infrared (FIR) and radio luminosities.
Since the FIR/radio derived SFRs are not affected by extinction, and
simple corrections to reconcile the UV and H$\alpha$ measurement with these
do not fully account for the discrepancies, a more sophisticated correction
may be required. Recent results suggest that at least part of the solution
may be a form of extinction which increases with increasing SFR (or luminosity,
given the common assumption that SFR is proportional to luminosity).
We present an analysis of the effects of a dust reddening {\em dependent on
star formation rate\/} applied to estimators of SFR. We show (1) that the
discrepancies between H$\alpha$ and FIR/radio SFR estimates
may be explained by such an effect, and we present an iterative method for
applying the correction; and (2) UV-based estimates of SFR are harder to
reconcile with FIR/radio estimates using this method, although the extent of
the remaining discrepancy is less than for a non-SFR-dependent correction.
Particularly at high redshift, our understanding of extinction at UV
wavelengths may require a still more complex explanation.
\end{abstract}

\keywords{dust, extinction --- galaxies: evolution --- galaxies: starburst}

\section{Introduction}
\label{int}

In recent years a variety of observations and observational methods
have been used to contribute to our understanding of the global
star formation history of the universe.
It is generally agreed that $\rho^*$, the
comoving space density of the star formation rate (SFR) in galaxies,
rises by an order of magnitude between $z=0$ and $z=1$
\citep{Lil:96,Con:97,Hog:98,Flo:99,Haa:00}. Despite several observations
encompassing $1<z<2$ and $z>2$ however,
\citep{Mad:96,Con:97,Hug:98,Yan:99,Ste:99,Hop:00} it is still unclear whether
the evolution of $\rho^*$ reaches a peak around $z\approx1.5$ and decreases
significantly thereafter \citep[e.g.,][]{Mad:96}, or whether it stays
flat to much higher redshifts \citep[e.g.,][]{Ste:99,Haa:00}.

Some of this uncertainty may be related to the difficulty of making
corrections for extinction at ultra-violet (UV), optical and near-infrared
wavelengths. In the absence of extinction corrections there is a clear
discrepancy between values of $\rho^*$ estimated from observations at radio
or far-infrared (FIR) wavelengths and those estimated from H$\alpha$ and UV
measurements. Even when simple extinction corrections to the UV and H$\alpha$
measurements are made, however, systematic differences
persist between estimates of SFR or $\rho^*$ made at differing wavelengths,
This can be seen, for example, in the distribution of SFRs for
(1) individual objects in a study of local galaxies by
\citet[their Figure~1]{Cram:98}, and in a multiwavelength study of Selected
Area 57 (SA57) by \citet[their Figure~3]{Sul:01};
and (2) globally, in a diagram showing the redshift dependence of
$\rho^*$ by \citet[their Figure~10]{Haa:00}.

Several efforts have been made to address this concern. A detailed
investigation of some of these issues is presented in \citet{AS:00}, with
particular emphasis on obscuration at UV wavelengths. \citet{Cal:99} have
modelled the evolution with redshift of galaxy opacity for applying corrections
to UV-derived estimates of $\rho^*$. \citet{Meu:99} have refined a technique
for correcting UV luminosities for attenuation using rest-frame luminosities
alone. This method is based on an observed correlation between the FIR/UV flux
ratio and $\beta$ the UV spectral slope. \citet{Gor:00} presents a more general
flux ratio method for establishing wavelength-specific attenuations for
individual galaxies, based on assumed SEDs.

We extend these investigations by first deriving an SFR-dependent
(or luminosity-dependent) attenuation from existing observations. We then
examine its efficacy by applying it to H$\alpha$ and UV estimates of SFR for
a large sample of local galaxies. The examination is subsequently expanded
to investigate the effects of the application to global SFR densities, derived
from H$\alpha$ or UV measurements, spanning a broad redshift range,

In Section~\ref{firjust} we comment on the use of FIR and 1.4\,GHz
luminosities as estimators of SFR. Section~\ref{meth} describes the
formulation of an SFR-dependent reddening parametrisation. This is
then applied to correct estimates of SFR for a sample of local
galaxies in Section~\ref{apply}, and subsequently to estimates of
$\rho^*$ spanning a broad redshift range in Section~\ref{sfrhist}.
Section~\ref{empirical} presents a brief investigation of the empirical
relation between SFR$_{\rm UV}$ and intrinsic SFR, and its effects when
used blindly to correct estimates of $\rho^*_{\rm UV}$ at all redshifts.

Throughout this paper we have used SFR calibrations defined consistently
to account for the mass range $0.1<M<100\,M_{\odot}$ assuming
a Salpeter IMF. We adopt the H$\alpha$ and UV SFR calibrations
given by \citet{Ken:98}, shown in Table~\ref{sfrcal}. Our assumed FIR
and 1.4\,GHz SFR calibrations are also given in this Table, and are
explained in more detail below.
We assume $H_0=75\,$km\,s$^{-1}$Mpc$^{-1}$ and $q_0=0.5$, and have
converted previously published estimates of SFR to our assumed calibrations
and this cosmology, where necessary, to ensure consistency.

\begin{deluxetable}{ll}
\tablewidth{0pt}
\tablecaption{Assumed luminosity-to-SFR calibrations.
 \label{sfrcal}}
\tablehead{
\colhead{Indicator} & \colhead{Calibration} \\
}
\startdata
UV & SFR$_{\rm UV}=L_{\rm UV}/7.14\times10^{20}\,{\rm W\,Hz^{-1}}$ \\
H$\alpha$ & SFR$_{\rm H\alpha}=L_{\rm H\alpha}/1.27\times10^{34}\,{\rm W}$ \\
FIR & SFR$_{\rm FIR}=L_{\rm FIR}/5.81\times10^{9} L_{\odot}$\\
1.4\,GHz & SFR$_{\rm 1.4GHz} = L_{\rm 1.4GHz}/8.4\times10^{20} {\rm W\,Hz}^{-1}$ \\
\enddata
\end{deluxetable}

\section{FIR and Radio estimates of SFR}
\label{firjust}

The intrinsic SFR of a galaxy cannot be estimated directly from
observations at UV or optical wavelengths because of the uncertainties
caused by the presence of unknown amounts of obscuring dust. Indicators
of SFR at longer wavelengths, at which dust is transparent,
are sought as an alternative. In the past few decades FIR luminosity
has been used extensively as an estimator of current star formation rate
in galaxies, and throughout this paper we have adopted the calibration given
by \citet{Ken:98}, which is shown in Table~\ref{sfrcal}. 
While FIR luminosity is expected to be an excellent SFR tracer 
for strong, compact, dusty starbursts, the situation is less clear when
the disks of normal galaxies are being examined. The uncertainties in
using FIR luminosity as an SFR estimator for normal galaxies are described
by \citet{Ken:98}, and while FIR luminosity in late-type star-forming galaxies
appears to correlate well with other estimators of SFR, this is not
generally the case for early-type spirals and ellipticals.

At still longer wavelengths, populations of star-forming galaxies have
been detected in increasingly sensitive radio surveys
\citep[e.g.][]{Wind:84,Wind:85,Oort:87,Hop:99}. As well as supporting
the well-established radio-FIR correlation for disk or
star-forming galaxies \citep[which extends over more than 4 orders of
magnitude and is commonly accepted to be a product of star formation
processes, see review by][]{Con:92}, estimates of SFR from measurements of
1.4\,GHz radio luminosity are now becoming more common
\citep[e.g.][]{Con:92,Cram:98,Haa:00}. We have adopted the 1.4\,GHz SFR
calibration as given by \citet{Haa:00},
SFR$_{\rm 1.4GHz} = Q (L_{1.4}/4.6\times10^{21} {\rm W\,Hz}^{-1})$,
with $Q=5.5$ the factor accounting for the
inclusion of stars less massive than 5\,$M_{\odot}$ (see Table~\ref{sfrcal}).

As a result, the use of radio or FIR luminosities as estimators of SFR
appear to be justified for samples of galaxies which lie on the radio-FIR
correlation. Admittedly, there are still concerns over the details of the
physical processes which produce this correlation \citep{Con:92}. While
the currently favoured explanation are processes related to star formation,
some radio-quiet quasars also appear to follow the correlation \citep{Cram:92},
as do Seyfert galaxies lacking compact radio cores 
\citep[although with greater scatter in this case]{Roy:98}.
The scatter in the radio-FIR correlation for disk galaxies, while smaller,
is also significant and its origins remain unknown.
Despite these concerns, long-wavelength estimators of SFR provide a valuable
tool which avoids the effects of obscuration, thus allowing useful
insights into the properties of star formation in various galaxy populations.

\section{An SFR-dependent reddening}
\label{meth}

Star forming galaxies in the local universe have been shown to exhibit
a correlation between obscuration and FIR luminosity, as seen for example in
\citet{Wang:96}. This implies that the obscuration in
a galaxy is related to its SFR. This same general effect can also be
seen in other recent results, e.g., \citet{AS:00}, particularly their
Figure~11, \citet{Cal:95}, and \citet{Sul:01}.

\citet{Wang:96} demonstrate a trend between FIR luminosity ($L_{\rm FIR}$) and
the inverse of the Balmer decrement for a small population of normal
Hubble types as well as some Markarian starbursts. A similarly small sample
of nuclear starburst and blue compact galaxies analysed by \citet{Cal:95}
show a clear trend between colour excess, (or UV spectral index), and
$L_{\rm FIR}$. Both trends are in the sense of increasing obscuration with
increasing $L_{\rm FIR}$. The scatter in these trends is notable, however,
and may be related to intrinsic differences within galaxy populations, as
well as between the different populations analysed.

\begin{figure*}
\centerline{\rotatebox{-90}{\includegraphics[width=8cm]{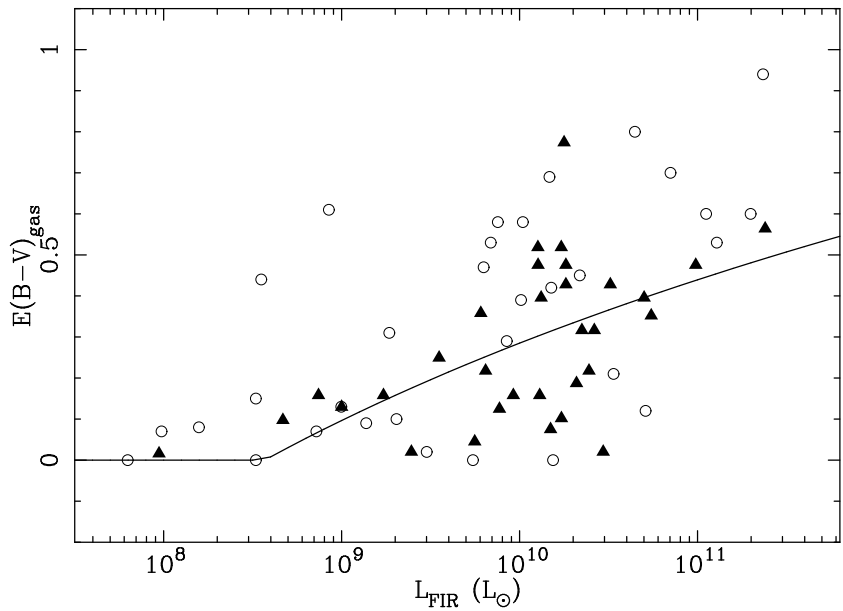}}}
\caption{The colour excess $E(B-V)$ as a function of FIR luminosity.
Solid triangles: data from \citet[their Figure 10 with Balmer decrement
converted to colour excess]{Wang:96}; Open circles: starburst and blue
compact galaxies from \citet{Cal:95}. The solid line shows the relation
between colour excess and FIR luminosity derived from combining
Equations \ref{bdecvsfir} and \ref{dust}.
 \label{ebvsfir}}
\end{figure*}

Now, since FIR luminosity for these objects can be treated as an estimator
of a galaxy's {\em intrinsic\/} star formation rate, this suggests that 
these relations can be recast in terms of obscuration (colour excess,
for example) and SFR$_{\rm FIR}$.
Using the data of \citet{Wang:96} we perform a least squares
fit to the Balmer decrement (not its inverse) and
$\log(L_{\rm FIR})$ to derive
\begin{equation}
\frac{F_{H\alpha}}{F_{H\beta}} = 0.797
  \log\left(\frac{L_{\rm FIR}}{L_{\odot}}\right) - 3.952.
 \label{bdecvsfir}
\end{equation}
The Pearson correlation coefficient for this fit is 0.6. \citet{Sul:01}
independently find a relation consistent with this result, from
observed Balmer decrements for their UV (200\,nm) selected-sample. 
$E(B-V)$, the colour excess appropriate for nebular emission lines,
is then calculated using the Balmer decrement \citep{Cal:96}\footnote{
We use the reddening curve $k(\lambda)=A(\lambda)/E(B-V)$ 
for star-forming systems formulated by \citet{Cal:00}:
\begin{eqnarray}
k(\lambda) & = & 2.659(-1.857 + 1.040/\lambda)+4.05 \nonumber \\
  & & \hspace{1.5cm} (0.63\,\mu m \le \lambda \le 2.20\,\mu m) \nonumber \\
           & = & 2.659(-2.156 + 1.509/\lambda - 0.198/\lambda^2 \nonumber \\
           &   & + 0.011/\lambda^3)+4.05 \nonumber \\
  & & \hspace{1.5cm} (0.12\,\mu m \le \lambda \le 0.63\,\mu m) \nonumber
\end{eqnarray}
}:
\begin{equation}
\label{dust}
E(B-V)_{\rm gas} = \frac{\log(R_{\rm obs}/R_{\rm int})}
  {0.4[k(\lambda_{H\beta})-k(\lambda_{H\alpha})]}.
\end{equation}
Here $R_{\rm obs}$ and $R_{\rm int}$(=2.88 for Case B recombination)
are the observed and intrinsic values for the Balmer decrement.
We can combine Equations \ref{bdecvsfir} and \ref{dust} to form a direct
relationship between colour excess and FIR luminosity. Figure~\ref{ebvsfir}
shows the resulting relation, along with the data used to derive
Equation~\ref{bdecvsfir}, \citep[modified from Figure~10 of][]{Wang:96}, and
data taken from the independent sample of starburst and blue compact galaxies
by \citet{Cal:95}. Since a negative colour excess is unphysical, we assume
that luminosities which give rise to $E(B-V)<0$ from this relation correspond
to zero attenuation, and this is seen in the flattening of the line
in Figure~\ref{ebvsfir}. Below $L_{\rm FIR}\approx4\times10^{10}\,L_{\odot}$
the data from \citet{Cal:95} appear to be consistent with the trend seen
in the \citet{Wang:96} data, but at higher luminosities even greater colour
excesses are seen. This suggests that for high-SFR galaxies in the local
universe, the following analysis may be somewhat conservative.

The relation between colour excess and FIR luminosity can now be transformed
to one between colour excess and SFR, using our adopted calibration for
SFR$_{\rm FIR}$, and we then find the colour excess appropriate to
a given {\em intrinsic\/} SFR:
\begin{equation}
\label{crux}
E(B-V)_{\rm gas} = 1.965 \log\left[\frac{0.797\log({\rm SFR_{FIR}}) + 3.834}
 {2.88}\right].
\end{equation}

This, then, forms the crux of an SFR-dependent reddening relation. Given
the intrinsic SFR for a galaxy (which may be derived from radio as well as
FIR luminosities, for example), and a reddening curve, the attenuation to be
applied at all wavelengths can be calculated.
Using the standard formulation \citep[e.g.][]{Cal:00,Cal:97} the
intrinsic luminosity, $L_i(\lambda)$, is related to the observed value,
$L_o(\lambda)$, by:
\begin{equation}
\label{standard}
L_i(\lambda)=L_o(\lambda) 10^{0.4 E(B-V) k(\lambda)},
\end{equation}
and since all the assumed SFR calibrations are directly proportional
to luminosity, this can be considered to be a relation between
intrinsic and observed SFR.

Additionally, and most importantly, since Equation~\ref{crux} is monotonic
even a measurement of the {\em attenuated\/} SFR can be used
to estimate the amount of reddening. This is done by substituting
Equation~\ref{crux} into Equation~\ref{standard} (cast in terms of SFR
rather than luminosity, since the SFR calibration constant cancels)
and solving numerically the resulting transcendental equation, which
for the wavelength of H$\alpha$ can be written:
\begin{eqnarray}
\label{hatransc}
\log(SFR_i) & = & \log(SFR_{o \rm (H\alpha)}) + 2.614\times \nonumber \\
   & & \log\left[\frac{0.797\log({\rm SFR_i}) + 3.834}{2.88}\right]. \nonumber\\
\end{eqnarray}
Hence, by assuming such a form for an SFR-dependent obscuration,
even observations of a single H$\alpha$ emission line (or other
star-formation-sensitive line, such as [OII]) may be used to
estimate attenuation and intrinsic SFR.

Now the colour excess appropriate for the stellar continuum,
$E(B-V)_{\rm star}$, is related to that for nebular gas emission lines by
\begin{equation}
\label{gasstar}
E(B-V)_{\rm star}=0.44 E(B-V)_{\rm gas},
\end{equation}
\citep[e.g.][]{Cal:00,Cal:97}. As a result it should in principle be possible
to apply the same procedure to derive appropriate corrections for
UV-continuum estimates of SFR. Combining Equations~\ref{gasstar}, \ref{crux}
and \ref{standard} gives the following relation for UV wavelengths:
\begin{eqnarray}
\label{uvtransc}
\log(SFR_i) & = & \log(SFR_{o \rm (UV)}) + X(\lambda_{\rm UV})\times \nonumber \\
   & & \log\left[\frac{0.797\log({\rm SFR_i}) + 3.834}{2.88}\right], \nonumber\\
\end{eqnarray}
where $X(365\,{\rm nm})=2.061$, $X(280\,{\rm nm})=2.512$ and
$X(150\,{\rm nm})=3.574$. Recent results from
\cite{AS:00} suggest, however, that observed (attenuated) UV luminosity
at short wavelengths ($\lambda\approx160$\,nm) is a very poor indicator
of SFR. They find that while galaxies with high SFRs
(and intrinsic UV luminosities) show increased levels of obscuration
\citep[see Figure~11 of][]{AS:00} consistent with our primary assumption,
the obscuration is such as to attenuate the observed UV luminosities to a
remarkably constant level for a broad range of
extinctions \citep[see Figure~17 of][]{AS:00}. This results in a
situation where Equation~\ref{uvtransc} effectively becomes
degenerate, and a knowledge of only the observed UV luminosity
is insufficient to regain information about the extent of the obscuration
and the corresponding intrinsic luminosity or SFR. In contrast with this
result, Figure~3 from
\citet{Sul:01}, comparing SFR$_{\rm UV}$ derived from observations at
$\lambda\approx200$\,nm to SFR$_{\rm 1.4GHz}$, does not show the expected
degeneracy, suggesting that perhaps the effect is stronger at shorter
wavelengths. As we later show, we also find no such degeneracy in U-band
derived SFRs, supporting this suggestion. The above method is thus
explored below with regard to UV luminosities using the relations already
derived, but placing emphasis on wavelengths longer than 160\,nm which appear
less likely to show this degenerate effect.

\section{Corrections to local SFR estimates}
\label{apply}

\begin{figure*}
\centerline{\rotatebox{-90}{\includegraphics[width=7cm]{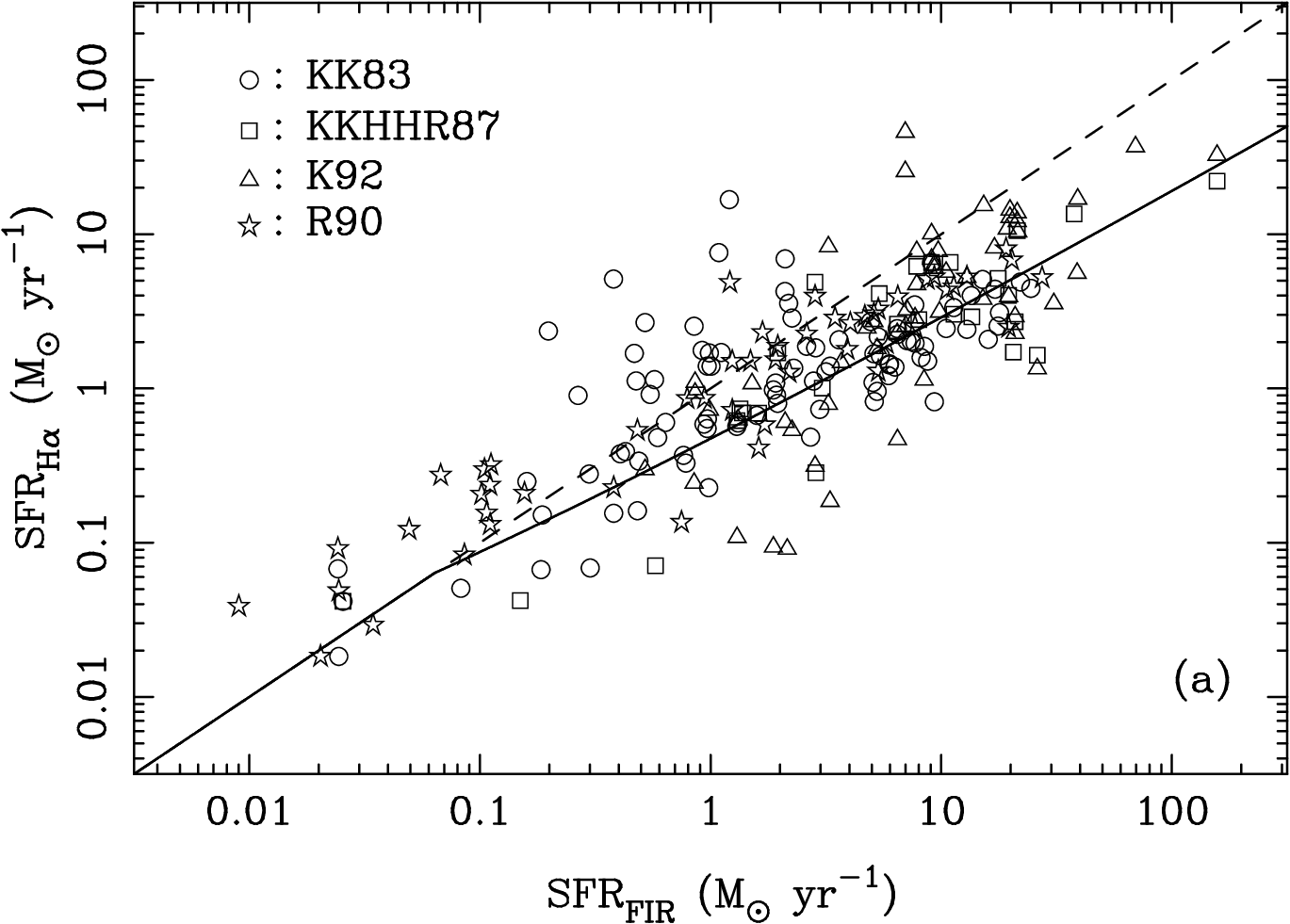}}
\hfill
\rotatebox{-90}{\includegraphics[width=7cm]{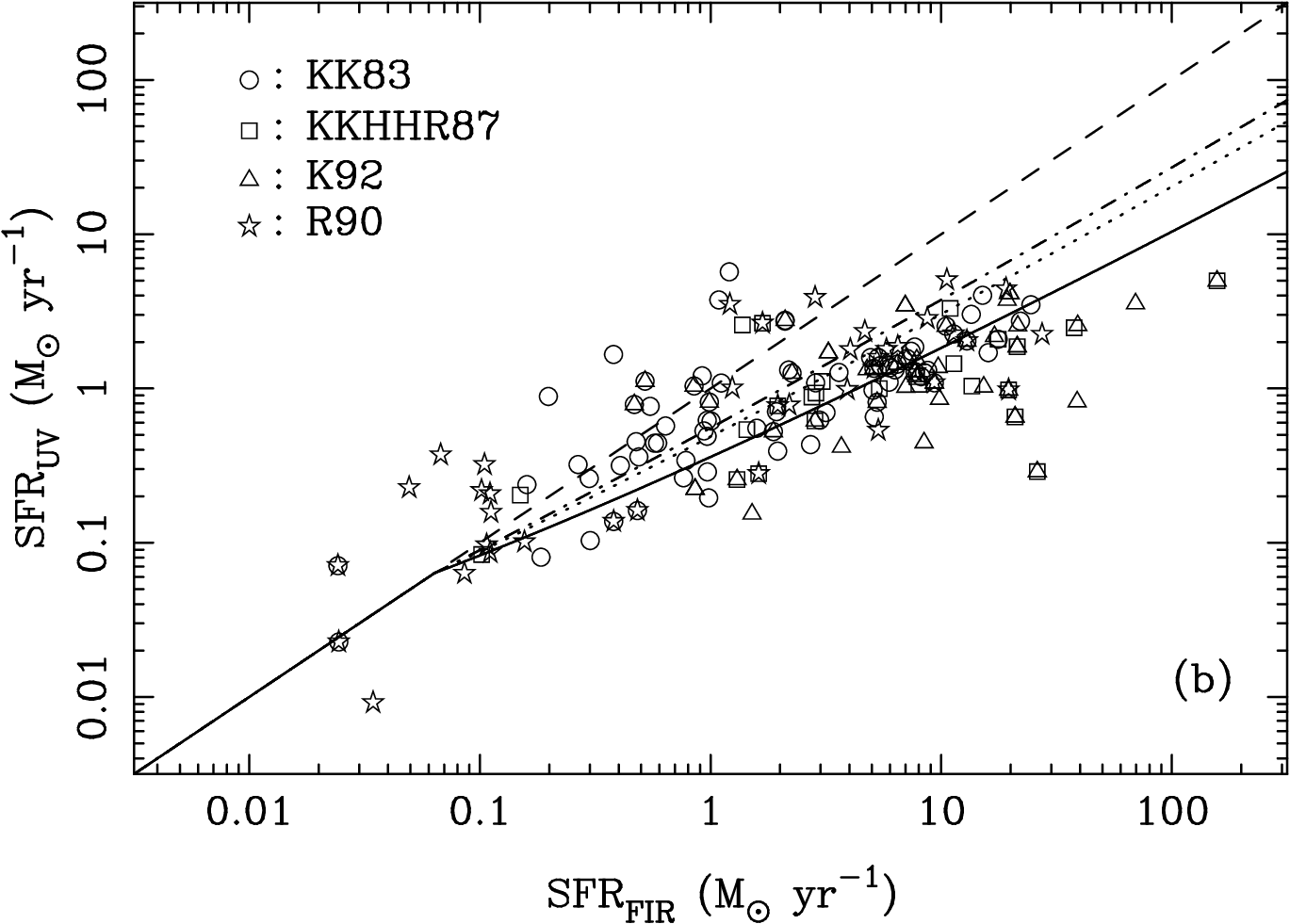}}}
\caption{(a) SFR$_{\rm H\alpha}$, and (b) SFR$_{\rm UV}$, both uncorrected
for extinction, compared to SFR$_{\rm FIR}$. The SFR$_{\rm FIR}$
values here come from measurements of the $60\,\mu$m luminosity, and the
SFR$_{\rm UV}$ values come from U-band (365\,nm) measurements. Both of these
are converted to SFR using the calibrations of \citet{Cram:98}, after
adjusting by a factor of 5.5 to account for stellar masses $<5\,M_{\odot}$.
Both calibrations are consistent with those of \citet{Ken:98}.
The solid line shows the SFR-dependent attenuation, calculated
using the prescription described in the text, for (a) H$\alpha$ and
(b) UV($\lambda=0.15\,\mu$m). The dashed line shows a one-to-one relationship.
The dot-dashed and dotted lines in (b) are for the attenuations valid
at $\lambda=0.365\,\mu$m (U-band) and $0.28\,\mu$m respectively. It can
easily be seen that the U-band attenuation in this formulation is not
sufficient to reproduce the observed trend. The different symbols mark
different sources for the data \citep[see also][]{Cram:98}:
KK83: \cite{KK:83}; KKHHR87: \cite{KH:87}; K92: \cite{Ken:92};
R90: \cite{Rom:90}.
 \label{extfir}}
\end{figure*}

\begin{figure*}
\centerline{\rotatebox{-90}{\includegraphics[width=7cm]{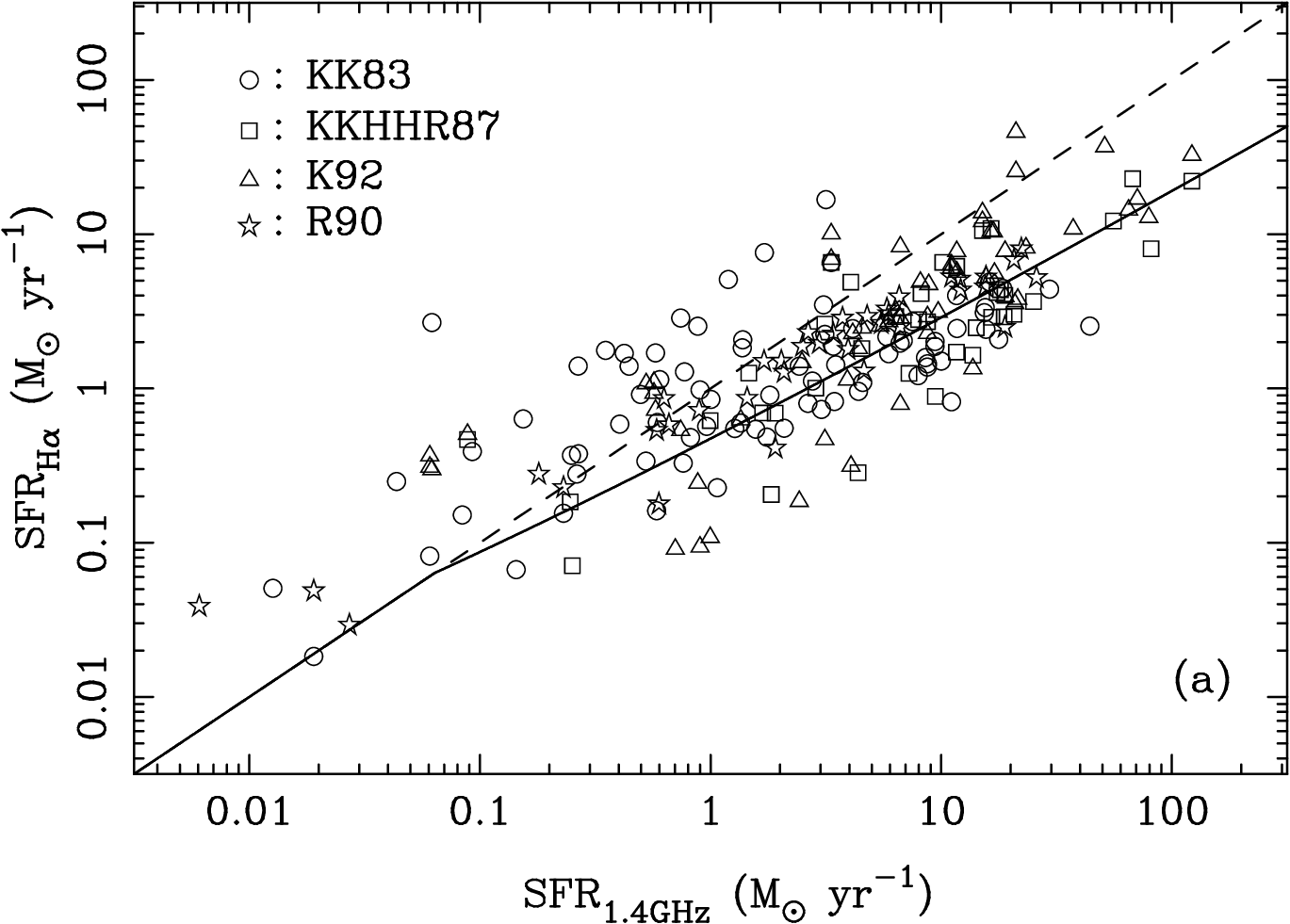}}
\hfill
\rotatebox{-90}{\includegraphics[width=7cm]{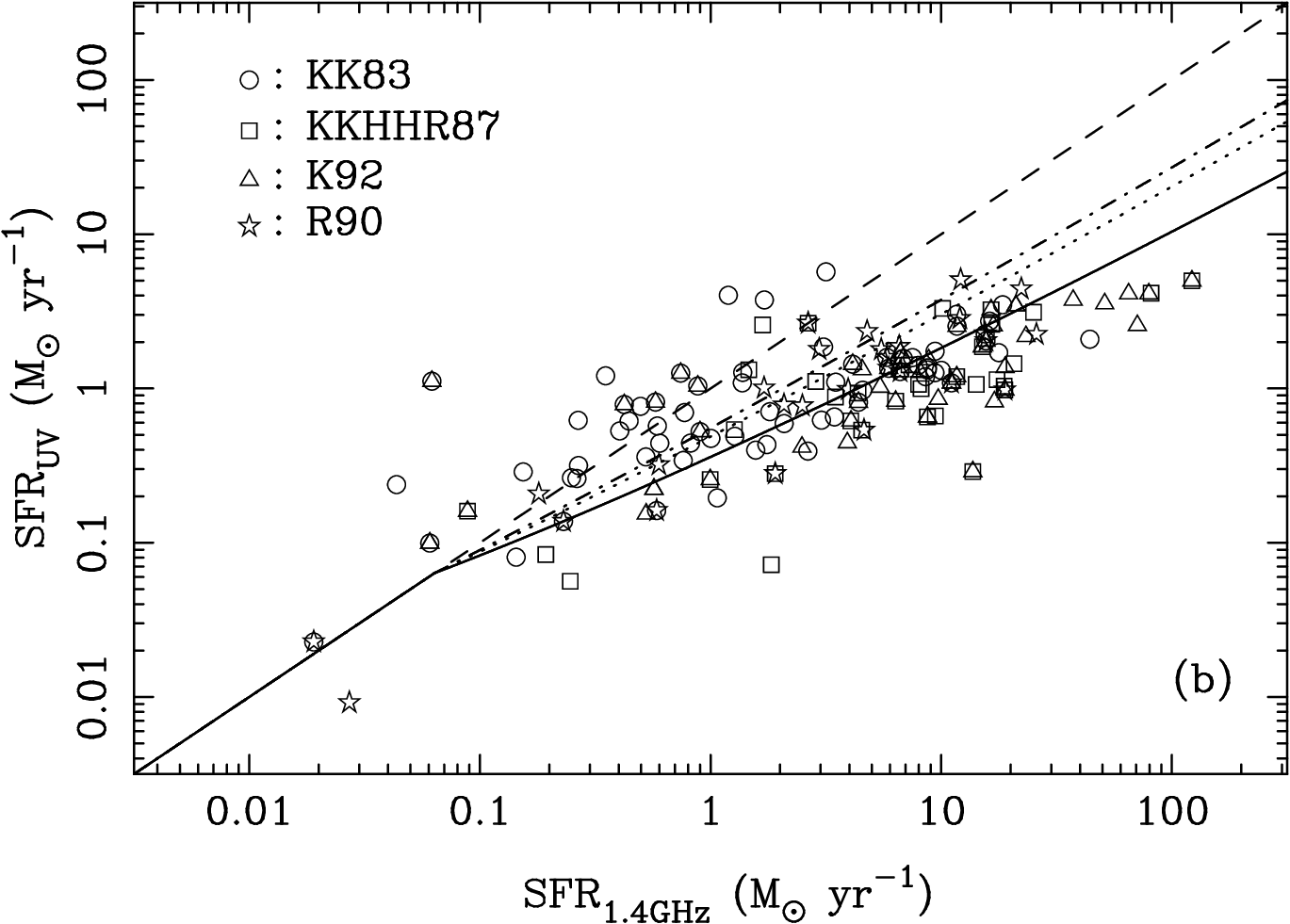}}}
\caption{As above, but relative to 1.4\,GHz derived SFRs instead of FIR,
emphasising the importance of 1.4\,GHz based estimates of SFR and
hence $\rho^*$.
 \label{extrad}}
\end{figure*}

The relationship between (obscured) H$\alpha$- and UV-based measures of SFR
and measurements of intrinsic SFR (for which we use
SFR$_{\rm FIR}$ or SFR$_{\rm 1.4GHz}$) are given by Equations~\ref{hatransc}
and \ref{uvtransc}. These relations are compared with observational
measurements for a sample of local galaxies in Figures~\ref{extfir} and
\ref{extrad}, using data selected from the compilation of \citet{Cram:98}.
Figure~\ref{extrad} emphasises the validity of the radio-FIR correlation for
this sample, and justifies the strong emphasis on 1.4\,GHz based estimates
of SFR (and subsequently $\rho^*$) in our analysis.
The ``knee" of the curves shown in Figures~\ref{extfir} and \ref{extrad}
marks the SFR at which the derived colour excess from Equation~\ref{crux}
goes to zero from higher values at higher SFRs.
For SFRs lower than this value we assume there is no attenuation of
the observed luminosity (and hence SFR).

The SFR-dependent reddening formulation seems to account well for the
observed discrepancy between SFR$_{\rm H\alpha}$ and SFR$_{\rm FIR}$ or
SFR$_{\rm 1.4GHz}$ (Figures~\ref{extfir}a and \ref{extrad}a). It is also
clear that, while this prescription may be useful for examining
the trends over large samples, the extent of
the intrinsic scatter in the relevant correlations seen in the observations,
is quite large, up to two orders of magnitude in some cases. Any detailed
analysis of individual galaxies should thus be treated with caution, and
quoted uncertainties in any derived results should reflect this.

The attenuation of the UV-derived SFRs, however, appears to be
less effectively modelled. As shown by the dot-dashed lines in
Figures~\ref{extfir}(b) and \ref{extrad}(b) the attenuation
implied at U-band by Equation~\ref{uvtransc} is insufficient to reproduce
the observed SFR$_{\rm UV}$. The observations of local galaxies imply
even greater levels of obscuration, and the degeneracy in observed UV
luminosities with respect to intrinsic UV luminosity found by \citet{AS:00}
possibly suggest greater levels still. The trend seen by \citet{AS:00}
could be illustrated in Figures~\ref{extfir}(b) and \ref{extrad}(b) by a
horizontal line at SFR$_{\rm UV} \approx 1\,M_{\odot}\,$yr$^{-1}$, for
intrinsic SFRs $\gtrsim 1\,M_{\odot}\,$yr$^{-1}$. This would intercept the
solid line in these Figures at a value of intrinsic
SFR $\approx5\,M_{\odot}\,$yr$^{-1}$, (less for the longer UV wavelength
models) and the implied obscuration would be greater than given by
Equation~\ref{uvtransc} (for $\lambda=150\,$nm) for objects with larger
intrinsic SFRs. Again, knowledge of only the observed UV luminosity
would be insufficient to establish the extent of the obscuration in
the degenerate case.

To establish whether or not the trend seen in the local U-band measurements
is consistent with the results of \citet{AS:00}, we have applied several
robust regression algorithms \citep{Iso:90} to the data of
Figures~\ref{extfir}(b) and \ref{extrad}(b). We find positive, non-zero
slopes in all cases, even when considering only data having values of
intrinsic SFR $\gtrsim 1\,M_{\odot}\,$yr$^{-1}$ (the
``flattest" portion of these diagrams). Our best estimate of the slope, from
the ordinary least squares bisector \citep{Iso:90}, is $0.7\pm0.07$ (for
logarithmic values of SFR), with a Pearson correlation coefficient of 0.5.
This may be compared with an approximate slope of 0.68 (for $0.1 <
$SFR$_{\rm FIR} < 10$) for the relation predicted by Equation~\ref{uvtransc}
at 150\,nm ({\em not\/} U-band). This result will be made use of in
Section~\ref{empirical}. While it is possible
that some incompleteness in the data has artificially skewed the slope
to positive values, it is also possible that there is a real trend in the
U-band data which is absent in the shorter wavelength (160\,nm) data. This
suggestion is not unreasonable, as the level of obscuration expected
at U-band (365\,nm) is less than at 160\,nm for a given intrinsic
luminosity, or SFR. It is also consistent with the non-degenerate
trend seen in Figure~3 of \citet{Sul:01} based on 200\,nm observations.
If this is the case, then it is possible that a method similar to the one
we analyse here may be useful for correcting UV observations at wavelengths
between U-band and 160\,nm. Indeed, it should be emphasised that the
empirical trends presented by the UV data (both in Figures~\ref{extfir}(b)
and \ref{extrad}(b), as well as Figure~3 of \citeauthor{Sul:01}
\citeyear{Sul:01}, and in \citeauthor{AS:00} \citeyear{AS:00})
are still evidence of a form of SFR-dependent obscuration, albeit one implying
even greater levels of attenuation than predicted by Equation~\ref{uvtransc}.

\section{Corrections to global SFR density}
\label{sfrhist}

We have so far considered the effects of a SFR-dependent obscuration
on samples of galaxies taken from the local universe. Now,
building on the conclusions of the previous Section, we consider how this
might affect the evolution of the global SFR density, given the
simplified assumption of no evolution in the properties or effects of dust.
In light of the shortfall
in the level of attenuation predicted by Equation~\ref{uvtransc}
we will treat the magnitude of the predicted corrections to $\rho^*_{\rm UV}$ 
as lower limits to the required correction, and examine the resulting
implications.

In applying the SFR-dependent reddening correction to samples at high
redshifts, we need to assume that the locally-derived trend between
obscuration and SFR is still valid. Obviously this may not be the case,
particularly since our understanding of galaxy evolution implies that
the metallicity of the inter-stellar medium increases with time. For
example, \citet{Ste:99} find a dearth of high-redshift galaxies with
$E(B-V) \gtrsim 0.3$. Extremely red galaxies, a class of objects defined
by $R-K>6$, may actually be a population of dusty star-forming galaxies at
high redshifts \citep[e.g.][]{Cim:99,And:00}, however, so the extent of
obscuration at high redshifts is still unclear
\citep[but see also the discussion of such objects in][]{AS:00}.
In the absence of more extensive evolutionary constraints, we begin with
this simple assumption and anticipate that (1) evolutionary effects may be
able to be incorporated as more data becomes available, and (2) remaining
discrepancies will serve to direct attention to those wavelengths and
redshifts deserving of further investigation.

\subsection{Applying the correction}

We present a compilation of uncorrected global SFR density ($\rho^*$)
measurements in Figure~\ref{sfd1}. This emphasises the magnitude of the
attenuation corrections required to reconcile the various estimates at
different wavelengths.
In addition to the UV surveys \citep{Con:97,Tre:98,Ste:99,Sul:00} and
H$\alpha$ surveys \citep{Gal:95,TM:98,Yan:99,Hop:00}, we
also show results from FIR observations \citep{Row:97,Flo:99} and
1.4\,GHz radio surveys \citep{Haa:00,Ser:00,Con:89}. All measurements have
been converted to our assumed cosmology and SFR calibrations for consistency.

To correct existing UV and H$\alpha$ estimates of
$\rho^*$ using the SFR-dependent reddening formulation described above
requires a knowledge not just of $\rho^*$ itself, but of the appropriate
H$\alpha$ or UV luminosity function, since the corrections to be
applied are SFR (and hence luminosity) dependent. We have initially
considered our derived relations, applying the corrections derived from
Equations~\ref{hatransc} and \ref{uvtransc} for a number of published
H$\alpha$ and UV surveys, sampling a broad redshift range. The results are
shown in Figure~\ref{sfd2}.

The local 1.4\,GHz measurement from \citet{Ser:00} of
$\rho^*_{1.4}=0.031\pm0.007\,M_{\odot}$yr$^{-1}$Mpc$^{-3}$
(after converting to $H_0=75$ and using our assumed SFR$_{\rm 1.4GHz}$
calibration), derived from a measurement of the local 1.4\,GHz LF of
Revised Shapley-Ames spiral galaxies, is consistent with the value of
$\rho^*_{1.4}=0.036\,M_{\odot}$yr$^{-1}$Mpc$^{-3}$ obtained by
integrating the local 1.4\,GHz LF for spiral galaxies from \citet{Con:89}.

Also shown in Figure~\ref{sfd2} are two models for the evolution of $\rho^*$.
The solid line is derived from a fit to the evolving 1.4\,GHz LF for
star-forming galaxies, measured by \citet{Haa:00} and based on the local LF
of \citet{Con:89}, and is anchored at $z=0$ by that point. The dashed line
is a model for the FIR luminosity density derived by \citet{Gis:00} from
measurements of the cosmic far-infrared background. This model has been
converted to SFR density using the SFR calibration of FIR luminosity
given by \citet{Ken:98}.

\subsection{Discussion}
\label{disc}

The effectiveness of correcting estimates of $\rho^*$ by assuming an
SFR-dependent obscuration is discussed separately below for three
broad redshift regimes.

\subsubsection{$0 < z \lesssim 0.3$}
\label{z1}

For redshifts out to $z\approx0.3$ applying
the SFR-dependent correction to (1) the uncorrected UV LFs of \citet{Sul:00}
and \citet{Tre:98}, and
(2) the N[{\sc ii}] and aperture corrected (but not extinction corrected)
H$\alpha$ LF of \citet{TM:98} all give consistent results.
The magnitude of the correction for $\rho^*_{\rm UV}$ in this redshift range
is a factor of about 3. While this correction may need to be considered
a lower limit the true value is probably not much greater than this, since
both corrected $\rho^*_{\rm UV}$ estimates and the corrected
$\rho^*_{\rm H\alpha}$ estimate are consistent with
the trend in the radio-based $\rho^*$ measurements spanning this
redshift range (see also Section~\ref{empirical}).

The one remaining discrepancy is a factor of $2-3$ difference between the
1.4\,GHz and H$\alpha$ based estimates of $\rho^*$ at $z \approx 0$. The
results of \citet{Gal:95} are shown in Figure~\ref{sfd2} with the reddening
correction originally applied by those authors. \citet{Mob:99} have also
derived a local H$\alpha$ SFR density which is consistent with this
measurement. The SFR-dependent formulation results in a correction similar to
the $\approx1$ magnitude often assumed for this local population, and is
unlikely to increase by more than a few tens of percent the value of
\citet{Gal:95}. (At higher redshifts the evolving H$\alpha$ LF provides greater
numbers of higher luminosity systems, which results in larger corrections
to $\rho^*_{\rm H\alpha}$.)

\citet{Ser:00} address this discrepancy by suggesting
that a significant fraction of star formation occurs in the cores of
giant molecular clouds, which would imply much greater obscuration than
derived from a Balmer decrement reddening correction assuming a simple
dust screen. Regardless of the mechanism, however, the obscuration must be
such that it explains the observed trends of Figures~\ref{extfir} and
\ref{extrad}. Hence, at least in the local universe, the correction
cannot be much greater than that which we examine in this analysis.
At higher redshifts, perhaps, the amount of attenuation could be greater,
although the trend of Figure~\ref{extfir}(a) is sufficient
to explain the discrepancy in $\rho^*_{\rm H\alpha}$ at $z \approx 1$
(see Section~\ref{z2}).

An alternative explanation for this discrepancy questions the calibration
of SFR$_{\rm 1.4GHz}$. This is ultimately based on the calibration of
non-thermal radio luminosity, $L_N$, to supernova rate, $\nu_{SN}$, in our
own galaxy \citep{Con:92}. If the escape of cosmic ray electrons from our
galaxy is significant, lowering the observed $L_N$ for a given SFR, then the
derived calibration constant will be larger than had we seen all the
radio emission. Hence, in external systems where we may see more, even most,
of the radio emission produced by the star-formation, the derived SFR will be
too great. \citet{Con:92} argues against this scenario, however, stating that
significant variations in the ratio $L_N/\nu_{SN}$ would violate the observed
radio-FIR correlation.

A third possibility is suggested by preliminary estimates of the
local H$\alpha$ luminosity density from the KPNO International Spectroscopic
Survey \citep[KISS,][]{Gro:97,Gro:99,Sal:00}. This survey finds a local
H$\alpha$ luminosity density somewhat higher than that measured
by \citet{Gal:95}.

\subsubsection{$0.3 \lesssim z \lesssim 2.0$}
\label{z2}

\begin{figure*}
\centerline{\rotatebox{-90}{\includegraphics[width=7cm]{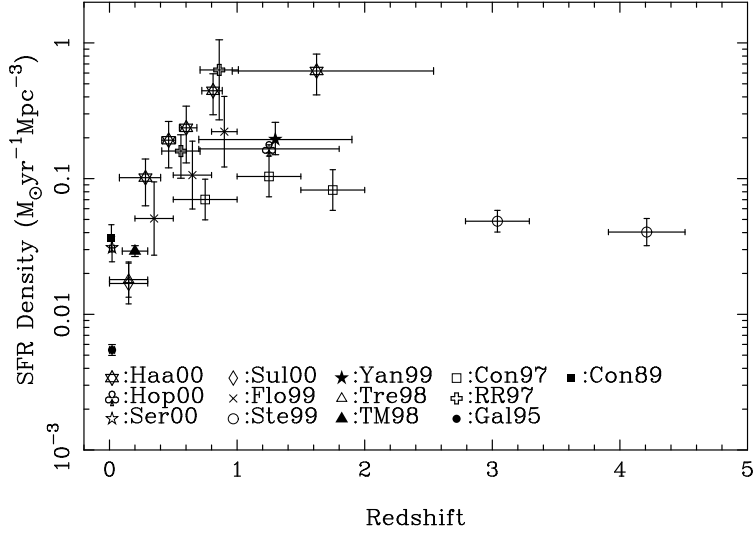}}}
\caption{SFR density ($\rho^*$) as a function of redshift. This diagram is
a compilation of SFR densities derived from various existing sources,
with no reddening corrections applied to any data. The reddening-corrected
measurement of \citet{Gal:95} has been artificially ``reddened" by 1
magnitude for comparison with other uncorrected data in this
diagram (solid circle). References in diagram are as
follows, along with the origin of the $\rho^*$ estimate:
Haa00: \cite{Haa:00} (1.4\,GHz); Hop00: \cite{Hop:00} (H$\alpha$);
Ser00: \cite{Ser:00} (1.4\,GHz);
Sul00: \cite{Sul:00} (UV); Flo99: \cite{Flo:99} (FIR);
Ste99: \cite{Ste:99} (UV); Yan99: \cite{Yan:99} (H$\alpha$);
Tre98: \cite{Tre:98} (UV); TM98: \cite{TM:98} (H$\alpha$);
Con97: \cite{Con:97} (UV); RR97: \cite{Row:97} (FIR);
Gal95: \cite{Gal:95} (H$\alpha$);
Con89: \cite{Con:89} (1.4\,GHz).
 \label{sfd1}}
\end{figure*}

\begin{figure*}
\centerline{\rotatebox{-90}{\includegraphics[width=7cm]{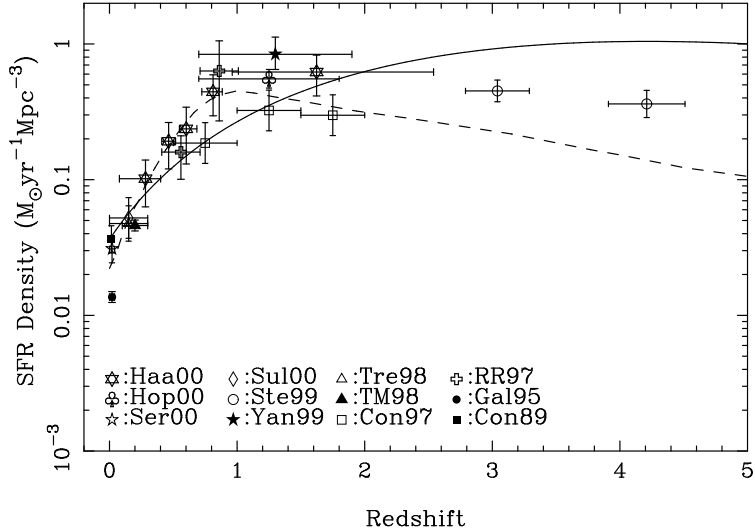}}}
\caption{SFR density ($\rho^*$) as a function of redshift. This diagram shows
the same data as in Figure~\protect\ref{sfd1} with the H$\alpha$ and UV
based measurements corrected using the SFR-dependent reddening prescription.
The FIR measurements of \citet{Flo:99} have been omitted in this
diagram for clarity.
The solid curve comes from the evolving 1.4\,GHz radio luminosity function for
star-forming galaxies derived by \citet{Haa:00}. The dashed curve is
a model derived from the cosmic far-infrared background by \citet{Gis:00}.
References are as for Figure~\protect\ref{sfd1}.
 \label{sfd2}}
\end{figure*}

Within this redshift regime there is a discrepancy between estimates
of $\rho^*_{\rm FIR}$ from different analyses. For
$0.3 \lesssim z \lesssim 1.0$, the FIR-based values of \citet{Row:97} are
comparable to the radio-based results of \citet{Haa:00}, but those of
\citet{Flo:99} lie a factor of $\approx1.5-2$ lower (Figure~\ref{sfd1}).
Possible sources for this discrepancy have been discussed by \citet{Flo:99},
who emphasise that their analysis covers a much larger area than that of
\citet{Row:97}, and uses different methods for fitting galaxy spectral
energy distributions.

\begin{figure*}
\centerline{\rotatebox{-90}{\includegraphics[width=7cm]{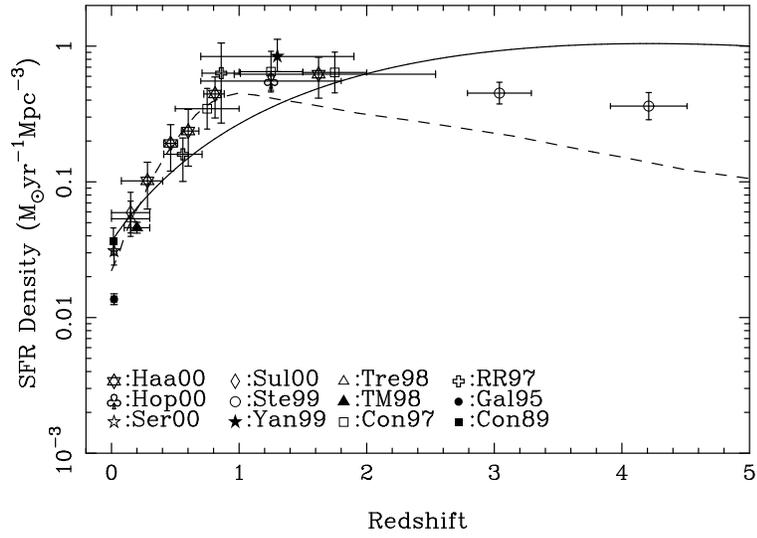}}}
\caption{SFR density ($\rho^*$) as a function of redshift. This diagram is
identical to Figure~\protect\ref{sfd2} except that the $\rho^*_{\rm UV}$ values
have all been corrected using the SFR-dependent reddening with
$k(\lambda)=k(0.15\,\mu$m), to examine the effects of applying the
locally observed empirical relationship between SFR$_{\rm UV}$ and
SFR$_{\rm FIR}$ to all redshifts.
 \label{sfd3}}
\end{figure*}

From $0.6\lesssim z<2.0$, the application of the SFR-dependent reddening to
H$\alpha$ LFs \citep{Yan:99,Hop:00} results in SFR densities which are
consistent with the observational radio-derived values, and the FIR-derived
values of \citet{Row:97}. This is highly encouraging, and suggests that 
the form of SFR-dependent reddening described by Equation~\ref{hatransc}
may be applicable as far out as $z \approx 2$. It also implies little
evolution in the form and extent of obscuration, at least in
a global, population-averaged sense. Obviously details for individual
galaxies will vary significantly, as evidenced by the large scatter
in Figures~\ref{extfir} and \ref{extrad}. The integration of the
luminosity function to derive estimates of $\rho^*$ has the effect of
averaging over this scatter, minimising its effect. The resulting integral
measure is thus more robust to the variations in individual galaxies,
and reflects instead the general trend such as that modelled by
Equation~\ref{hatransc}. A further observational estimate of
$\rho^*_{\rm H\alpha}$ at $z\approx0.9$ \citep[not shown in
Figures~\ref{sfd1}-\ref{sfd3}]{Gla:99} may also
be consistent with these results, as they comment that their value
is consistent with the ``high" estimate of \cite{Row:97} if they
apply the Calzetti prescription for dust attenuation.

The same treatment with the UV LFs of \citet{Con:97} give a factor of
$\approx 4$ corrections to $\rho^*_{\rm UV}$. The corrected $\rho^*_{\rm UV}$
still lies a factor of $\approx1.5-2$ lower than $\rho^*_{\rm FIR}$,
$\rho^*_{\rm 1.4GHz}$ or the corrected $\rho^*_{\rm H\alpha}$
(Figure~\ref{sfd2}). This is not unexpected, and is consistent with
the interpretation of Equation~\ref{uvtransc} as providing lower limits
for the corrected $\rho^*_{\rm UV}$.

It is possible, on the other hand, that lower metallicities and
obscurations at higher redshifts for galaxies of a given luminosity
imply some evolution in the calibration of SFR$_{\rm UV}$. Such an
effect would be in the sense that a given {\em observed\/} UV luminosity at
higher redshifts implies {\em lower\/} levels of SFR. This would give a
calibration coefficient which decreases with redshift, reducing the observed
$\rho^*_{\rm UV}$, and the correction for any obscuration would also
decrease with redshift. If this effect is large enough,
it could exacerbate the discrepancy between $\rho^*_{\rm UV}$ and
the 1.4\,GHz and FIR estimates. This extreme scenario
would be inconsistent with the results of the H$\alpha$ corrections in this
redshift regime, which suggest that any such evolution in the
level of obscuration is probably small.

\subsubsection{$z \gtrsim 2.0$}
\label{z3}

At the high redshift end, $z\approx3-4$, the UV LFs of \citet{Ste:99}
have also been re-examined assuming an SFR-dependent reddening. Despite the
large uncertainties associated with extrapolating Equation~\ref{uvtransc}
to high redshifts, and the expectation that in its present form the
correction is likely to be underestimated, it is encouraging to see that
the results lie midway between the predictions of the two models shown.
The corrected estimates using Equation~\ref{uvtransc} are bracketed by the
predictions from the model evolving radio LF and the FIR background model.
They are also consistent with the lower limit
derived from sub-mm SCUBA observations of the Hubble Deep Field \citep{Hug:98},
$\rho^*>0.16\,M_{\odot}$yr$^{-1}$Mpc$^{-3}$ in our cosmology.
The correction derived from Equation~\ref{uvtransc} approaches a factor
of 10 at $z\approx4$, and may {\em still\/} need to be considered a lower
limit to the true correction given the results of \citet{AS:00}. The extremely
large uncertainties associated with extrapolating the locally-derived
SFR-dependent reddening models to such high redshifts, though, dictate
the use of caution when interpreting these results. It is still
interesting to note that if these corrections are indeed
underestimates and the true values need to be higher, they would lie
closer to the extrapolations from the 1.4\,GHz luminosity functions
estimated by \citet{Haa:00}, almost an order of magnitude above
the estimates from the FIR background.

\section{An empirical investigation}
\label{empirical}

The inadequacy of Equation~\ref{uvtransc} in fully explaining the observed
levels of obscuration prompts two questions. The obvious one
relates to the origin of the obscuration, and is discussed briefly
below. The second is: ``Independent of the mechanism
producing the obscuration, is the local {\em empirical\/} relation
between SFR$_{\rm UV}$ and intrinsic SFR sufficient to explain the
discrepancies in $\rho^*$ at higher redshifts?"
We examine this by applying the empirical relationship for local galaxies
observed in Figures~\ref{extfir}(b) and \ref{extrad}(b) to $\rho^*_{\rm UV}$
measurements at all redshifts.

For this simple analysis we treat the relationship between
SFR$_{\rm UV}$ and SFR$_{\rm FIR}$ (or SFR$_{\rm 1.4GHz}$) described by the
$0.15\,\mu$m attenuation (the solid curve in these Figures) as a mathematically
convenient description of this empirical relation. We also assume this
relationship is valid at shorter wavelengths \citep[where the attenuation
may be still greater, given the results of][]{AS:00},
but particularly at $0.28\,\mu$m, where the measurements from \citet{Con:97}
were made. Applying the SFR-dependent correction to all the
$\rho^*_{\rm UV}$ points based on this empirical result gives the
diagram shown in Figure~\ref{sfd3}. Good agreement is now seen between all
UV, H$\alpha$ and radio-derived values for $\rho^*$. The effectiveness
of this application is dominated by the effective correction to the
data of \citet{Con:97}, since the corrections at low-redshift are not
modified greatly (due to the LF being dominated by lower-luminosity sources).
The primary conclusion here is that the empirical relation between
local U-band estimates of SFR$_{\rm UV}$ and intrinsic SFR is sufficient
to account for obscuration in 280\,nm observations at $z \approx 1.5$.

Two possible scenarios to explain the shortcomings of Equation~\ref{uvtransc}
are discussed in this light. First, dust in star forming galaxies may
be characterised by a reddening curve even greyer than that of \citet{Cal:00}.
The situation may be explained if the attenuation at wavelengths between
$0.15 < \lambda \lesssim 0.4\,\mu$m is comparable to that observed
locally in U-band. One model producing this effect may be a heavily
dust-enshrouded starburst, optically thick below $0.4\,\mu$m, but having
``holes" which allow some portion of UV light to escape unattenuated.
This would result in a uniform ``blocking" of all emission below the optically
thick cutoff, and produce the required constant attenuation as a function
of (UV) wavelength. The effective reddening curve for this scenario
may, however, not be consistent with the results of the FIR dust
luminosity analysis presented in \citet{Cal:00}.
Alternatively, the dust attenuation remains as given by \citet{Cal:00}, but
there is some additional process which needs to be invoked to explain the
shortfall in the U-band derived SFRs seen in Figures~\ref{extfir}(b) and
\ref{extrad}(b). This may be the case, for example, if the relative
contribution of the old stellar population to the U-band luminosity is greater
at low SFR. (Luminosities at shorter UV wavelengths would have less
``contamination" from the old stars, so would be affected less or not at all
in this scenario.) This would imply that the U-band derived SFRs in low-SFR
systems are {\em over}-estimated, and need to be revised downward. Then,
combined with a suitable revision of the U-band/SFR calibration, this would
account for the discrepancies. There are still problems with this
suggestion as well, since low-SFR systems also tend to be low-luminosity
systems \citep[e.g.][their Figure~3]{Cram:98}, and this implies the
{\em relative\/} contribution from the old stellar population may not change
much with SFR.

In addition, whatever mechanism is invoked needs to be able to reproduce
the degenerate relationship between observed UV luminosity and obscuration
at shorter wavelengths found by \citet{AS:00}.
More investigation of the result described here is obviously still required.
Although the application of the locally observed empirical SFR$_{\rm UV}$ to
SFR$_{\rm FIR}$ relation to shorter wavelengths and higher redshifts may 
account for the $\rho^*_{\rm UV}$ discrepancies it does not explain their
physical origin. It does emphasise that the trend observed
locally in U-band is consistent with the discrepancies at much higher
redshifts and at shorter wavelengths.
This is consistent with the result for the H$\alpha$ correction
suggesting little net evolution, if any, in the level of obscuration
at least out to $z\approx2$.

\section{Conclusions}
\label{conc}

We have modelled an SFR-dependent attenuation by dust, characterised by
the Calzetti reddening curve. Corrections derived from this model have been
applied to a large sample of local galaxies, spanning a wide range in intrinsic
SFR, to examine the effects on local H$\alpha$- and UV-derived SFR estimates.
Subsequently, measurements of $\rho^*$ for a broad range of redshift have
also been corrected using this model.
It is clear from the observational data investigated here, as well as
the smaller samples of \citet{Wang:96} and \citet{Cal:95}, that some form
of SFR-dependent extinction is implied. The investigated prescription
appears sufficient to explain the general trend
between SFR$_{\rm H\alpha}$ and SFR$_{\rm 1.4GHz}$ or SFR$_{\rm FIR}$. It
also accounts for the discrepancies between estimates of
$\rho^*_{\rm H\alpha}$ and $\rho^*_{\rm 1.4GHz}$ or $\rho^*_{\rm FIR}$ out
to $z\approx2$, with the exclusion of the local $\rho^*_{\rm H\alpha}$ estimate
of \citet{Gal:95}. This is strongly suggestive of little or no evolution
in the extent or form of the obscuration at rest-frame H$\alpha$ wavelengths.
As a result, this may serve as a useful tool in making corrections for
obscuration in estimates of SFR$_{\rm H\alpha}$ or $\rho^*_{\rm H\alpha}$
in the absence of more direct methods (such as for the NICMOS grism surveys 
of \citet{Yan:99} and \citet{Hop:00}, and the recent SOFI/ISAAC results of
\citet{Moo:00}, which lack Balmer decrement information).

The examined prescription for correcting UV-based estimates of SFR and
$\rho^*$ was less effective. Although the predicted corrections reduce
the level of observed discrepancy with obscuration-free estimators, they
do not fully account for the observed levels of attenuation in local
U-band estimates of SFR$_{\rm UV}$. There may also be concerns that
a degenerate relation (at shorter wavelengths) would prevent application
of the iterative method used. The use of this method to estimate lower
limits for the corrections to $\rho^*_{\rm UV}$
is more encouraging, though, particularly for $z\lesssim0.3$ where the
magnitude of the required correction is unlikely to increase by much.
The situation becomes more complex at higher redshifts, and here
evolutionary effects, as well as a more complete understanding of
the obscuration in galaxies at UV-wavelengths, may need to be incorporated.

The empirical relationship observed at U-band for local galaxies between
SFR$_{\rm UV}$ and SFR$_{\rm FIR}$ or SFR$_{\rm 1.4GHz}$ was also
examined. It was used to correct high-redshift estimates of
$\rho^*_{\rm UV}$, resulting in values consistent with $\rho^*_{\rm 1.4GHz}$
and $\rho^*_{\rm FIR}$ for UV wavelengths longer than 160\,nm. This may
support the result from the H$\alpha$ analysis, suggesting minimal evolution
in the extent of dust obscuration. The mechanisms producing the obscuration
are still in question, and prompt further investigation of dust models and
the characteristics of dust on UV-wavelength radiation.

\acknowledgements

The authors wish to thank an anonymous referee for several highly
constructive comments, and AMH gratefully thanks Herv{\'e} Dole for
drawing his attention to the results of Gispert, Lagache \& Puget.
AMH and AJC acknowledge support provided by NASA through grant numbers
GO-07871.02-96A and NRA-98-03-LTSA-039 from the Space Telescope Science
Institute, which is operated by AURA, Inc., under NASA contract NAS5-26555.
This research has made use of the NASA/IPAC Extragalactic Database (NED)
which is operated by the Jet Propulsion Laboratory, California Institute of
Technology, under contract with the National Aeronautics and Space
Administration.

\end{document}